# Rejection of Power Supply Noise in Wheatstone Bridges: Application to Piezoresistive MEMS

El Mehdi Boujamaa, Yannick Soulie, Frederick Mailly, Laurent Latorre and Pascal Nouet

Laboratoire d'Informatique, de Robotique et de Microélectronique de Montpellier (LIRMM)
UMR CNRS 5506 - University Montpellier II – France
Contact: latorre@lirmm.fr

***Abstract*- This paper deals with the design of MEMS using piezoresistivity as transduction principle. It is demonstrated that when the sensor topology doesn't allow a perfect matching of strain gauges, the resolution is limited by the ability of the conditioning circuit (typically a Wheatstone bridge) to reject power supply noise. As this ability is strongly reduced when an offset voltage is present at the output of the bridge, the proposed architecture implements a feedback loop to control MOS transistors inserted in the Wheatstone bridge to compensate resistor mismatches. This feedback exhibits a very good offset cancellation and therefore a better resolution is achieved.**

## I. Introduction

During the last two decades, a considerable transformation of the "Sensor world" was achieved by the emergence of MEMS technologies. Even if transducers based on capacitive sensing are the most commonly used, piezoresistive sensors that are intrinsically less sensitive have a unique advantage of being fully compatible with CMOS technologies. As an example some of the authors already proposed both inertial and magnetic field sensors on the same CMOS die [1]. At the price of a simple post-process, it is thus possible to integrate on a single wafer, heterogeneous systems including one or several electro-mechanical sensors and their conditioning circuitry. The integration of conditioning electronic close to the sensing element, improves the overall performances. Indeed having the sensor and the conditioning circuitry on the same substrate avoid the sensor to be loaded by long cables and to prevent noise and interference pickup [2].

The principle of a piezoresistive sensor is to convert mechanical stress (or strain) into a resistance variation. The transduction is realised thanks to strain gauges designed with materials that have piezoresistive properties. Those gauges are resistances that we want to measure the value with a conditioning (biasing) circuit. In our study, we will focus on the Wheatstone bridge, which is the most common conditioning structure used for this type of sensors [3].

In the case of a perfectly balanced Wheatstone bridge, the four resistances are identical in the absence of signal and the power supply noise is totally rejected. Assuming that the differential voltage at the output of the bridge may be amplified without degradation of the signal to noise ratio, the resolution of the sensing system is then limited by the intrinsic thermal noise generated by resistors forming the Wheatstone bridge.

In this paper, a feedback conditioning architecture is proposed to reject the power supply noise and to maximise the resolution up to the intrinsic resolution of the Wheatstone bridge (SNR=1). This study is illustrated using a magnetic field sensor developed by our research team [4] but can be extended to any piezoresistive resonant sensors. In the next section, the sensor principle is introduced along with the basis of power supply noise rejection. The principle of the offset cancellation scheme is discussed and validated in section 3. Finally, the hardware implementation is presented in section 4 where the efficiency of the proposed circuit is evaluated.

## II. Sensor Principle and PSRR Background

The magnetic field sensor is based on a "U-shape" mechanical structure as shown on Figure1. It is obtained after wet anisotropic etching of the silicon bulk and is composed with the various backend layers of the standard CMOS process. Among them, oxide and nitride layers are the base materials for the frame, polysilicon and aluminium are respectively used to embed strain gauges and a coplanar coil. The later is flowed by a current *If* that interacts with the earth magnetic field and leads to a Lorenz force that bends the frame. The mechanical strain generated by the bending, translates into a change of gauge resistances placed closely to the anchor of the suspended "U-shape".

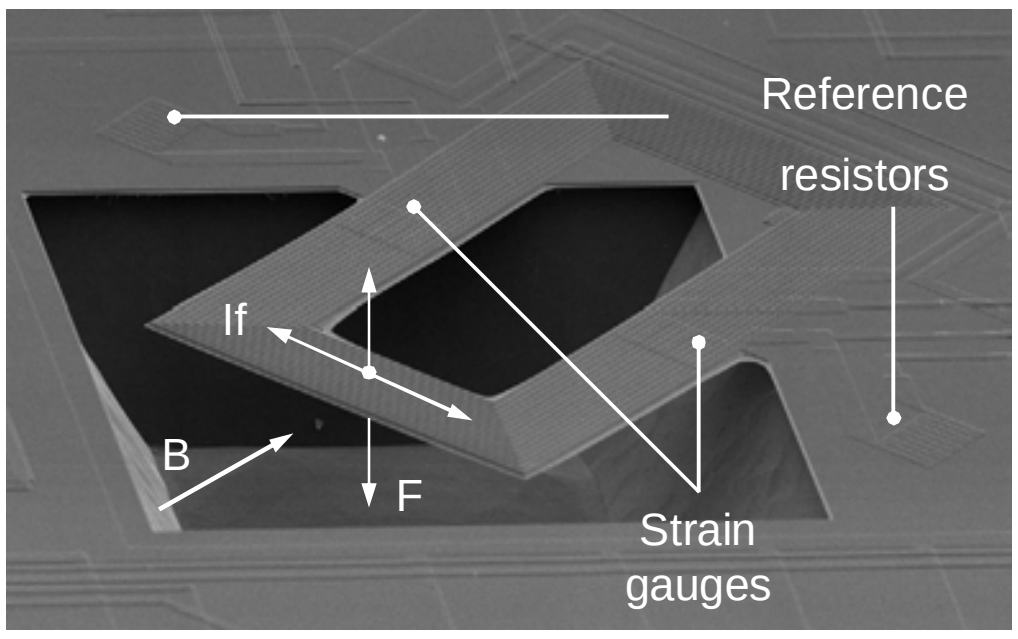

Fig. 1. The piezoresistive magnetic field sensor.

Both gauges (namely R1 and R4) are arranged in a Wheatstone bridge together with two reference resistors (namely R2 and R3) deposited over the bulk. Figure 2 illustrates this arrangement of gauges and reference resistors.

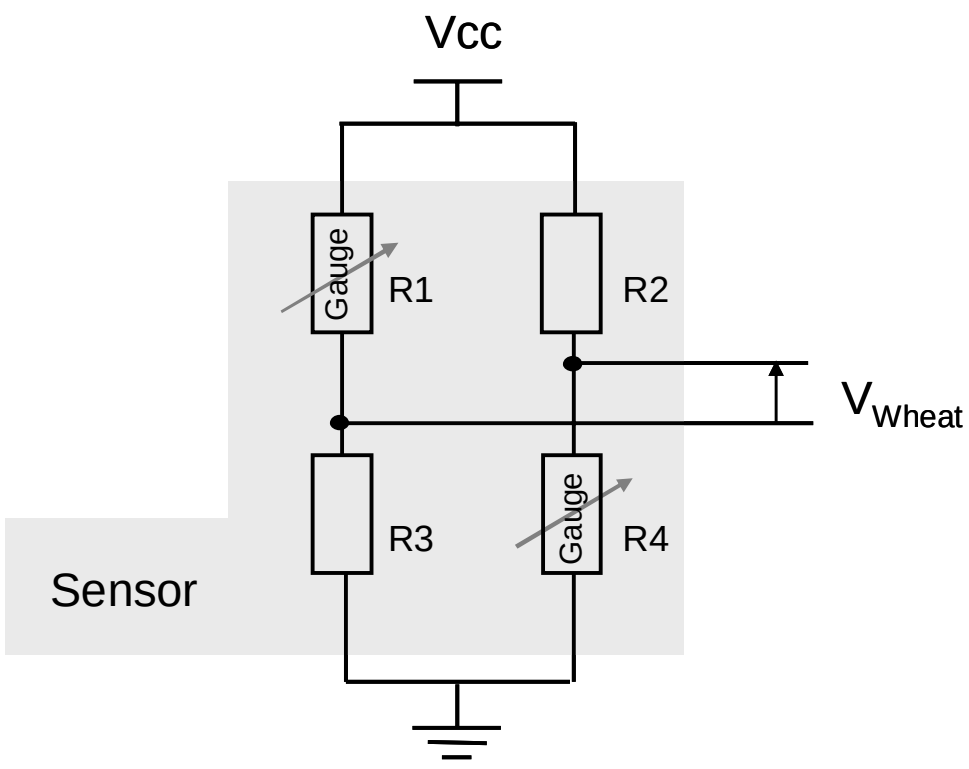

Fig. 2. Conditioning circuit for strain gauges.

The differential output voltage (VWheat) of the Wheatstone bridge writes:

$$V_{Wheat} = V_{cc} \cdot \frac{R_2 \cdot R_3 - R_1 \cdot R_4}{(R_1 + R_3)(R_2 + R_4)} \quad (1)$$

Thus when the bridge is balanced (R2.R3=R1.R4) the output voltage is null and therefore the power supply noise is perfectly rejected. In general, the ability of an electronic circuit to reject the power supply noise is characterised by the PSRR (standing for Power Supply Rejection Ratio). This figure is defined as the ratio between the gain w.r.t. the input signal and the gain w.r.t. a change of the power supply:

$$PSRR = 20\log\left(\frac{\partial V_{out}/\partial V_{in}}{\partial V_{out}/\partial V_{dd}}\right) \quad (2)$$

In our case, the frame is a second order mechanical system, with a weak damping or a high quality factor. The resonance phenomenon can be used to increase the sensor sensitivity. To make so, the coil current is set to an ac current at the cantilever resonant frequency (22 kHz in our case). The output voltage is then composed with a carrier at 22 kHz, modulated by the amplitude of the earth magnetic field. This signal is detected by an amplification circuitry composed with a Low Noise differential Amplifier (LNA), high pass filters, and differential amplifiers to reach a gain of about 25000 (figure 3).

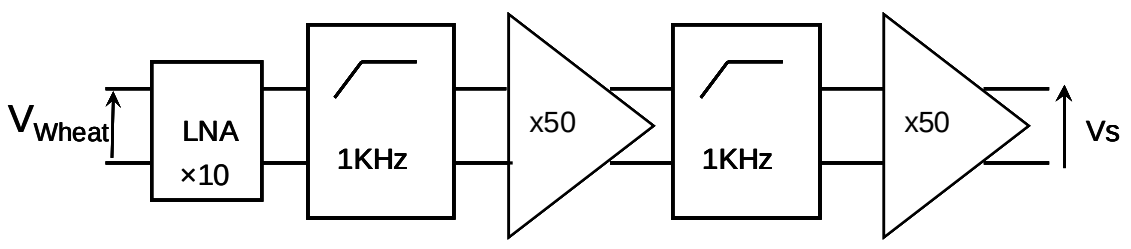

Fig. 3. Open loop amplification chain

The amplification chain was designed to reach an optimal performance in terms of intrinsic noise. The equivalent input noise equals to 6.4 nV/√Hz (at 22 kHz).

This noise level is equivalent to the intrinsic noise of the sensor, i.e. the thermal noise of bridge resistances (1kΩ at 300°K) thus maximising the resolution of the sensor.

However, the resolution for the complete system degrades due to resistance mismatches between gauges and reference resistors. In our case, the power supply noise is only attenuated by 49.5dB. In the literature it can be found that the power supply noise, introduced by a digital function in a circuit, can reach several hundreds of millivolts [5]. Therefore, making a realistic assumption that a 10µV/√Hz wide band (1MHz) noise affects the power supply, the resulting noise at the bridge output equals few dozens of nV/√Hz; far more than the intrinsic noise level. Finally, it comes that the resolution limitation is due to the Wheatstone bridge PSRR.

Improving the PSRR requires a better balanced bridge. Let's review the source of mismatches. On the one hand, the reference resistors are placed over the substrate whereas the gauges are embedded in the suspended frame. This leads to different internal stresses between reference resistances and gauges and thus to different initial value of resistances. Different orientations also induce a higher susceptibility to CMOS spreading. On the other hand, gauges are subject to self-heating. Indeed, because of the bridge biasing an increase in temperature of gauges is obtained. This is due to the fact that gauges are thermally isolated from the substrate. This phenomenon is even worse considering the cumulative effect of self-heating and power in the excitation coil [6]. Due to the temperature coefficient of the polysilicon, this leads to a new source of mismatch. Finally, there are so many reasons for having an unbalanced bridge and having a self-balanced bridge is not achievable simply by design.

## III. PSRR IMPROVEMENTS

As previously established, designing a self-balanced bridge is not an easy task. It is then necessary to implement a robust design strategy in order to ensure better bridge balance, to reduce bridge output offset and hence to reach a better rejection of the power supply noise.

The following principle has been studied: a voltage-controlled resistor is added in each branch of the bridge to balance it. The added resistor values are directly controlled by the amplified Wheatstone bridge differential output voltage. In figure 4, this principle is represented as a block diagram. For simplicity reasons, a half bridge is represented where $\beta$, $A$, $R_{DS}$ and $\Delta R$ are respectively the feedback coefficient (Ω/V), the gain of the voltage amplifier, the value of the controlled resistor (Ω), and the mismatch of the bridge.

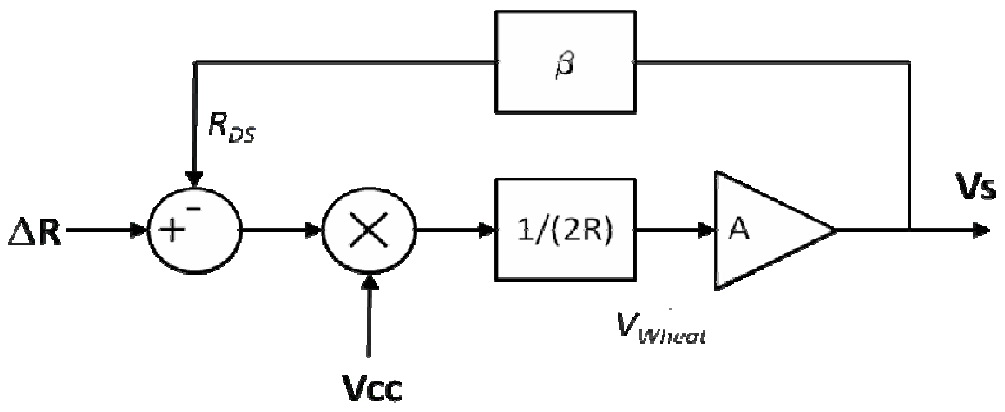

Fig. 4. Block diagram of the enhanced-PSRR architecture

From this diagram, the output voltage writes:

$$V_s = \Delta R \frac{\frac{A}{2R}Vcc}{1+\frac{A\beta}{2R}Vcc} \quad (3)$$

Partial derivations of this equation lead to the relative variation of the output voltage w.r.t. the power supply (noise) or the input signal (resistance variation):

$$\frac{\partial V_s}{\partial \Delta R} = \frac{\frac{A}{2R}Vcc}{1+\frac{A\beta}{2R}Vcc} \approx \frac{1}{\beta} \quad (4)$$

$$\frac{\partial V_s}{\partial Vcc} = \frac{A\cdot\Delta R}{2R\cdot\left(1+\frac{A\beta}{2R}Vcc\right)^2} \approx \frac{2R\cdot\Delta R}{A\beta^2\cdot Vcc^2} \quad (5)$$

Considering $A\beta/2R >> 1$, it is obvious that the output signal and the power supply induced noise are proportional respectively to $1/\beta$ and $\Delta R/A\beta^2$. In the following, the criterion used to evaluate the ability of the circuit to reject power supply noise is defined as:

$$PSRR^{-1} = 20\log\left(\frac{\partial V_S / \partial V_{CC}}{\partial V_S / \partial \Delta R}\right) = 20\log\left(\frac{2R.\Delta R}{A\beta V_{CC}^{\ 2}}\right) \quad (6)$$

For a perfectly balanced bridge, where *ΔR* is null, a perfect power supply noise rejection is achieved. Figure 5 illustrates the evolution of $PSRR^{-1}$ versus the input mismatch *ΔR*, for different values of the amplifier gain.

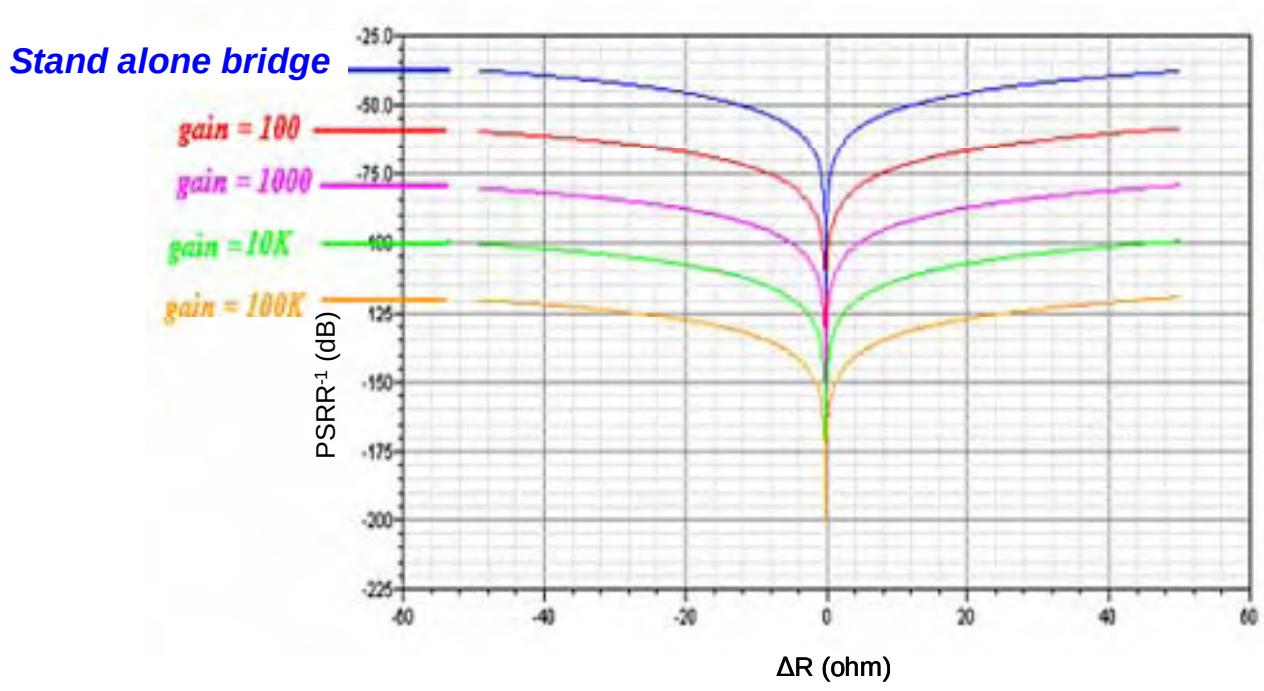

Fig. 5. PSSR-1 with and without feedback loop, for different values of the amplifier gain.

From these simulations, equation (6) is verified by demonstrating a 20 dB PSRR increase for each decade of gain. These simulations were performed using Cadence® tools. The implemented structure is illustrated figure 6, where the feedback resistors (MOS transistors), were described using a VERILOGA description language.

## IV. HARDWARE IMPLEMENTATION

A first implementation using PMOS transistors in triode region has been developed. In figure 6, a simple implementation is shown with two transistors used as voltage controlled resistances and a fully differential amplifier. Transistor sizing is trivial and won't be developed here.

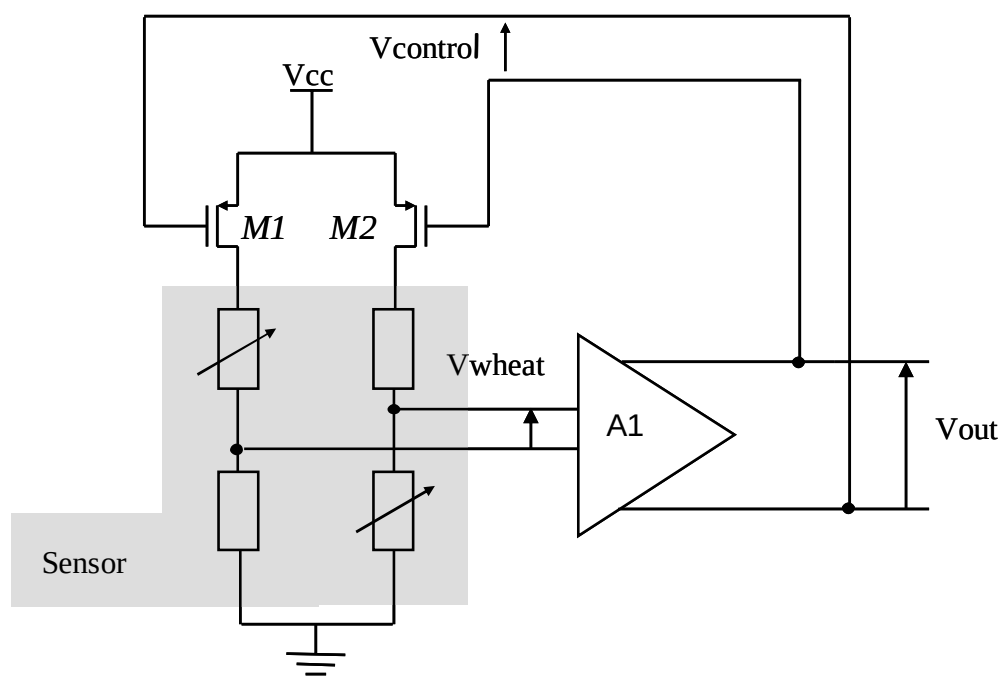

Fig. 6. Bridge balancing feedback with PMOS transistors.

MOS resistance is controlled by the gate-to-source voltage. The source being tied to Vcc, power supply noise leads to a $V_{GS}$ modulation and thus to a $R_{DS}$ variation. Both transistors finally act as common gate amplifiers for the power supply noise thus degrading the PSRR. It is worth noting that, even if inefficient for the power supply noise, this architecture rejects the ground noise. Simulations are reported on figure 7 where a 1 kHz resistance variation is applied to the gauge together with a 9 kHz noise on the ground. Results clearly show that the noise is completely rejected: the 9 kHz tone is not present on the differential outputs (curves b and c).

Similarly, the use of NMOS transistors placed between the ground and the bridge resistors will eliminate power supply noise while amplifying ground noise.

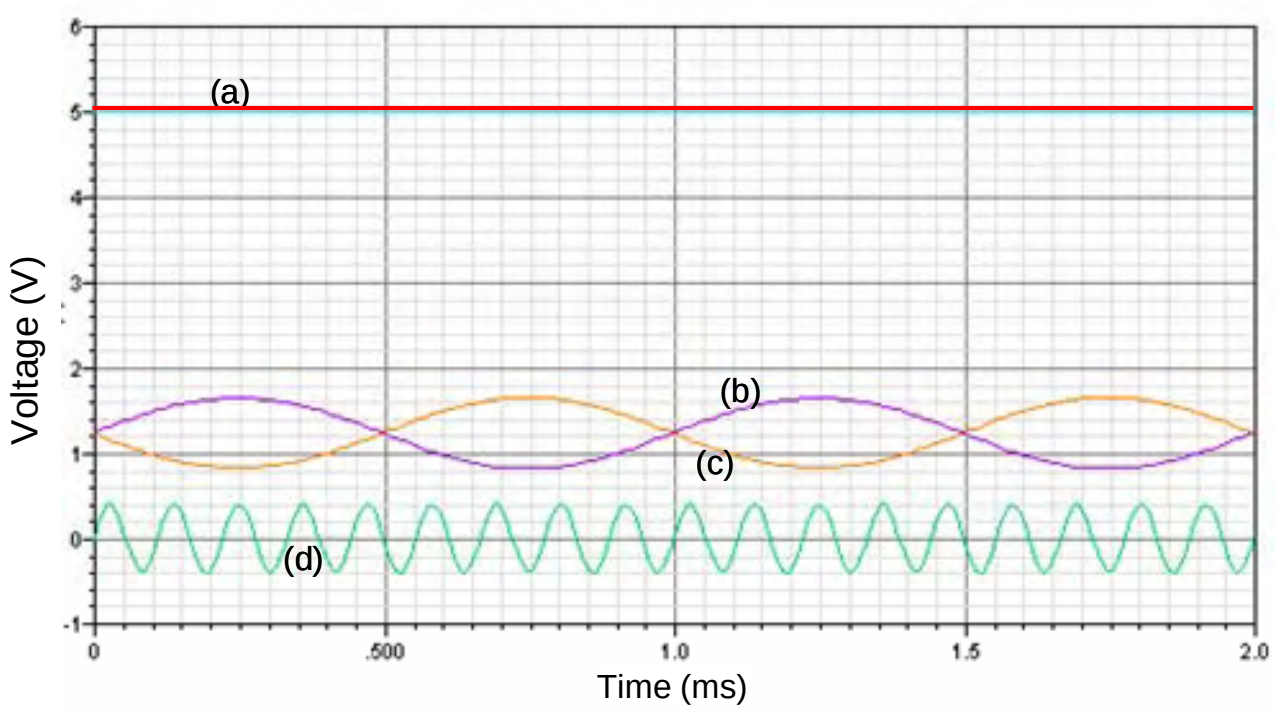

Fig. 7. Transient simulation in presence of both signal (@1kHz) and ground noise (@9kHz): a) Vcc, b) positive and c) negative differential outputs, d) Gnd.

A second implementation is depicted in figure 8 where an RC circuit has been added between each differential outputs of the amplifier and the power supply. From amplifier outputs we obtain a low pass filter that cancels only bridge static offset. From the power supply, a high pass filter is obtained and the gate-to-source voltage modulation is cancelled in the frequency range of the signal.

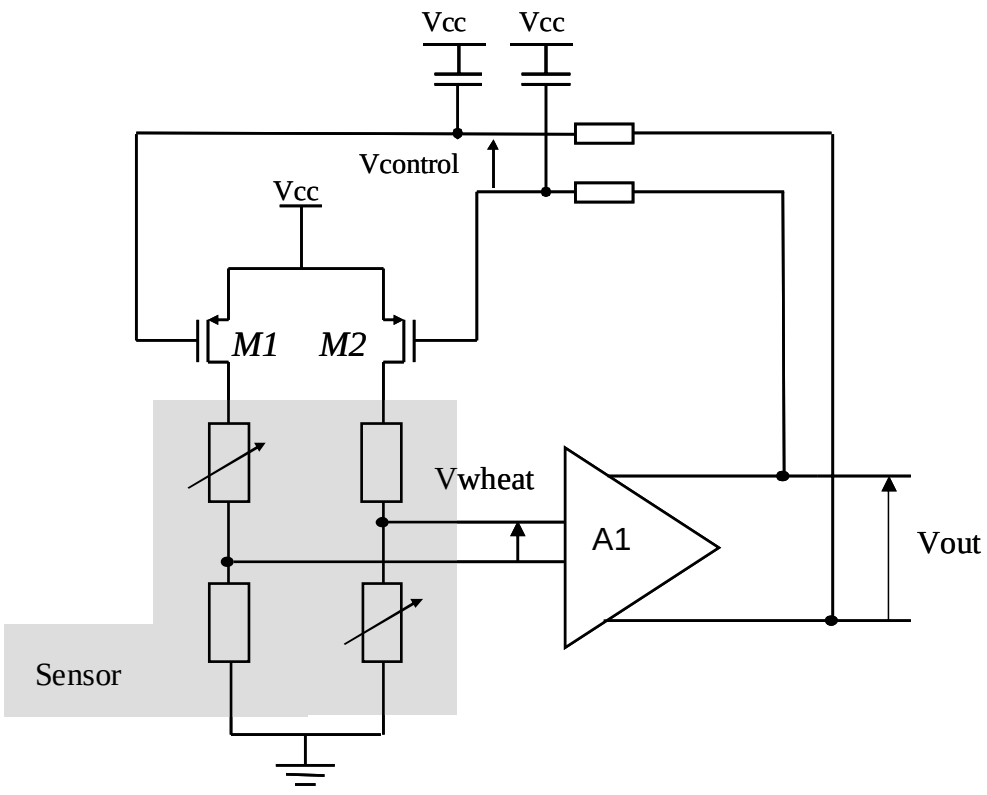

Fig. 8. Enhanced PSRR circuit for both Vcc and Gnd noise rejection.

Figure 9 illustrates the power supply noise rejection versus frequency. In a 1 KHz - 100 KHz frequency band, the compensated bridge exhibit a PSRR more than ten times higher than the stand alone Wheatstone bridge.

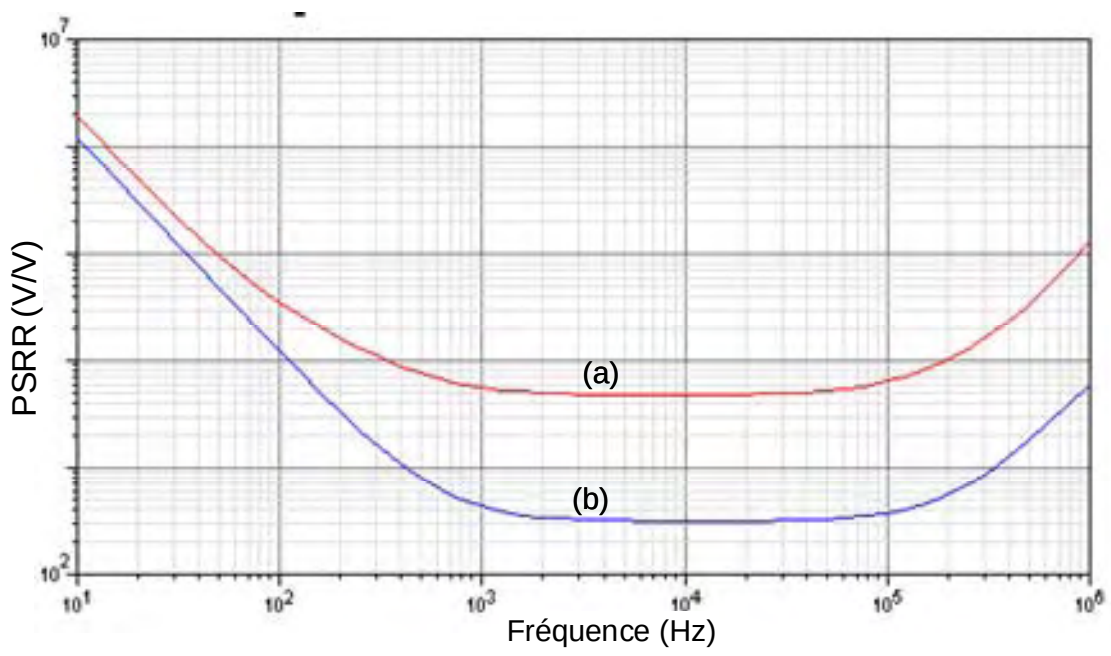

Fig. 9. PSRR simulation for a compensated (a) and stand alone (b) Wheatstone bridge.

Considering that the noise rejection architecture has been validated, we then characterize the overall performance of the proposed solution. Assuming feedback transistors are sized with a sufficient area to reduce their intrinsic noise, we have compared the open-loop architecture (figure 3) with the previously proposed one for the same overall gain in the range of 87 dB. It comes that, even if the PSRR is increased, the overall performance is reduced due to the differential amplifier noise. We then modified the differential amplifier to include the LNA of the open-loop architecture before the differential amplifier. Obtained results (last column in table 1) demonstrate a very good compromise with very high PSSR and a very good intrinsic noise level.

TABLE 1
OVERALL NOISE PERFORMANCE OF THE STUDIED ARCHITECTURES

| | **Open loop** | **Feedback loop (no LNA)** | **Feedback loop (with LNA)** |
|---|---|---|---|
| Gain | 87dB | 86.9dB | 85.9dB |
| PSRR | -49dB | -73dB | -84dB |
| Intrinsic electronic noise | 160 µV/√Hz | 390 µV/√Hz | 150 µV/√Hz |

## V. CONCLUSION

In this paper, we have proposed a feedback conditioning architecture to improve the power supply noise rejection ratio (PSRR) of a Wheatstone bridge. The proposed architecture is applicable to any resonant piezoresistive MEMS to improve the resolution. This architecture is of primary importance in the case of highly integrated MEMS where the power supply may be polluted by the switching of digital electronics.

Efficiency of the proposed solutions has been demonstrated using high-level simulations based on Matlab or VERILOGA languages to validate the principle and transistor level simulations for final implementation and overall performance evaluation.

The final architecture exhibits the same intrinsic noise level, the same sensitivity (gain) and a 35 dB PSRR improvement. It is then possible to calculate the resolution of the sensor using only the SNR ratio within gauges.